\documentclass[aps,  twocolumn, prd, superscriptaddress, nofootinbib,longbibliography]{revtex4-2}

\usepackage{amsmath, amssymb, bm}
\usepackage{graphicx}
\usepackage{epstopdf}
\usepackage{amsfonts}
\usepackage{amssymb}
\usepackage{amsbsy}
\usepackage{amsmath}
\usepackage{latexsym}
\usepackage{sansmath}
\usepackage{subfigure}
\usepackage{lipsum}
\usepackage{float}
\usepackage{cancel}

\usepackage{bm}
\usepackage{xcolor}

\usepackage{comment}
\usepackage{csquotes}				
\usepackage{physics}
\usepackage{accents}
\usepackage{appendix}

\usepackage{hyperref}
\usepackage{nameref}

\def\bea {\begin{eqnarray}}
\def\eea {\end{eqnarray}}
\def\nn {\nonumber}

\begin{document}

\title{Regular black holes and boson stars in semiclassical gravity}

\author{Irfan Javed} \email{ijaved@unb.ca}
\affiliation{Department of Mathematics and Statistics, University of New Brunswick, Fredericton, Canada}

\author{Viqar Husain} \email{vhusain@unb.ca}
\affiliation{Department of Mathematics and Statistics, University of New Brunswick, Fredericton, Canada}

\begin{abstract}
We use a Hamiltonian version of the semiclassical Einstein equation to study classical gravity coupled to a quantum scalar field with potential in spherical symmetry. The system is defined by effective constraints where the matter terms are replaced by their expectation values in a quantum state. For the static case, we find numerically that the resulting equations admit asymptotically flat, de Sitter, and anti-de Sitter boson star and regular black hole solutions. We also show that the Bardeen and Hayward nonsingular black hole proposals (and some generalizations thereof) are not solutions to our equations.
\end{abstract}

\maketitle

\section{Introduction}
\label{sec. 1}
Quantum gravity (QG) research has been ongoing for more than six decades with no consensus on a final theory \cite{carlip2015quantum, loll2019quantum, aastrup2012intersecting}. The goal in QG is to find a unifying framework for a dynamical theory of matter and geometry with quantum theory. A central problem in this endeavor is that standard quantum theory requires a fixed background geometry for its very definition, a feature that is not available if the metric is a dynamical degree of freedom.

A proposed intermediate ``approximation" for gaining insight into QG is the semiclassical Einstein equation
\bea
G_{ab}(g) = 8\pi G\ \langle\hat{T}_{ab}(g, \hat{\phi})\rangle_\Psi,
\label{scee}
\eea
where the stress-energy tensor is an operator that depends on the unknown ``semiclassical metric" $g$ and the expectation value is calculated in a specified quantum state $|\Psi\rangle$ of the matter field $\hat{\phi}$. This equation is a proposal for a coupled classical-quantum system interpreted in the Heisenberg picture, where the operator $\hat{T}_{ab}(g,\hat{\phi})$ can be defined. However, $\hat{T}_{ab}$ is not known \textit{a priori} because it depends on $g$, the unknown of the problem. This poses questions for the very definition of this equation. However, the equation can be interpreted in perturbation theory; for a discussion of various issues, see, e.g., \cite{isham1995structural, ford2005spacetime}. Furthermore, such a semiclassical approximation can be derived for bipartite quantum systems, but its range of validity is limited \cite{singh1989notes, husain2023motivating}.

These problems with the formulation of (\ref{scee}) may be avoided, at least to some extent, by considering a canonical formulation of semiclassical gravity. For general relativity, the classical canonical equations schematically take the form
\begin{equation}
\begin{aligned}
\dot{g} &= \{g, H[g, \phi; N]+C[g, \phi; N^{a}]\},\\
\dot{\phi} &= \{\phi, H[g, \phi; N]+C[g, \phi; N^{a}]\},
\end{aligned}
\label{canon}
\end{equation}
where $g, \phi$ respectively denote the canonical variables for geometry and matter; $N, N^{a}$ the lapse and shift functions; and $H$ and $C$ the Hamiltonian and diffeomorphism constraints.

We define canonical semiclassical equations by quantizing the matter field and replacing the matter terms in the constraints and the $\dot{g}$ evolution equations by their expectation values in a quantum state. This leads to ``effective" constraints $H_{\rm eff} = 0$ and $C_{\rm eff} = 0$ and $\dot{g}$ equations $\dot{g} = f(N, N^{a}, g, \bra{\Psi}\hat{\phi}\ket{\Psi})$, where $f(N, N^{a}, g, \phi)$ is the result of the $\dot{g}$ Poisson bracket in (\ref{canon}). The functional Schrodinger equation for the quantum state replaces the canonical evolution equations for matter in (\ref{canon}). Schematically, the semiclassical equations so defined are
\begin{equation}
\begin{aligned}
H_{\rm eff} &\equiv H_{G}+\bra{\Psi}\hat{H}_{M}\ket{\Psi} = 0,\\
C_{\rm eff} &\equiv C_{G}+\bra{\Psi}\hat{C}_{M}\ket{\Psi} = 0,\\
\dot{g} &= f(N, N^{a}, g, \bra{\Psi}\hat{\phi}\ket{\Psi}),\\
\iota\dot{\ket{\Psi}} &= \left(\hat{H}_{M}[g, \hat{\phi}; N]+\hat{C}_{M}[g, \hat{\phi}; N^{a}]\right)\ket{\Psi},
\end{aligned}
\label{CSCE}
\end{equation}
where $G$ denotes the gravity-only component of our constraints and $M$ the matter-only one. This system requires initial data $(g_0, \ket{\Psi}_0)$ that satisfy the effective constraints; the data are then evolved with the $\dot{g}, \dot{\ket{\psi}}$ equations. This is a canonical Schrodinger picture realization of semiclassical gravity; it has been applied to homogeneous cosmology \cite{husain2019semiclassical, husain2021quantum, husain2024semiclassical}.

In this work, we apply it to the gravity-scalar field model in spherical symmetry with a focus on static solutions, where the gravitational phase space variables are time independent (i.e., $\dot{g} = 0$) and the state $|\Psi\rangle$ evolves only by a phase (i.e., $\bra{\Psi}\hat{\phi}\ket{\Psi}$ are fixed in time). The system (\ref{CSCE}) then reduces to
\begin{equation}
\begin{aligned}
H_{\rm eff} &= 0,\quad C_{\rm eff} = 0,\\
0 &= f(N, N^{a}, g, \bra{\Psi}\hat{\phi}\ket{\Psi}).
\end{aligned}
\label{static1}
\end{equation}
As we shall see, the last of these equations consists of a set of coupled equations for the gravitational phase space variables in spherical symmetry, the lapse and shift, and the functions of the radial coordinate $r$ that represent expectation values of the scalar field operators. We solve these equations numerically with the appropriate conditions for asymptotic flatness, de Sitter (dS), and anti-de Sitter (AdS) cases and with the matter operator expectation value functions chosen to represent bounded energy densities. By specifying matter expectation values rather than states, no approximations are required to implement full backreaction.

Our approach stands in contrast to the usual approach to semiclassical gravity where the first step is the computation of $\langle\hat{T}_{ab}\rangle$ on a fixed and specified background \cite{birrell1984quantum}. It is also different from the related semiclassical self-consistent configurations (SSCs) approach \cite{diez2012towards}, for the latter requires explicit calculation of the quantum state $\ket{\Psi}$ as well, which we do not have to bother with while still considering full matter backreaction. In fact, our approach does away with all approximations that go into the calculation of $\langle\hat{T}_{ab}\rangle$ in SSCs, such as the ``large $N$" approximation. It further avoids approximations to $\langle\hat{T}_{ab}\rangle$ in scenarios other than SSCs too, for example, the Polyakov approximation \cite{kain2025quantum}.

The outline of the paper is as follows. Sec.~\ref{sec. 2} presents the derivation of the spherically symmetric semiclassical equations; Sec.~\ref{sec. 3} gives the boson star and black hole solutions; Sec.~\ref{sec. 4} gives a proof that the solutions we find are not the same as some other nonsingular black hole proposals; and Sec.~\ref{sec. 5} contains a summary and discussion of our results.

\section{Semiclassical equations in spherical symmetry}
\label{sec. 2}
The theory we consider is gravity coupled to a real scalar field with action (in $G = c = \hbar = 1$ units)
\begin{equation}
\begin{aligned}[b]
S =&\:\frac{1}{16\pi}\int d^{4}x\sqrt{-g}^{(4)}R-\frac{1}{8\pi}\int d^{4}x\sqrt{-g}\ (g^{\alpha\beta}\partial_{\alpha}\phi\partial_{\beta}\phi)\\
&\:-\frac{1}{4\pi}\int d^{4}x\sqrt{-g}\ V(\phi)
\end{aligned}
\label{action}
\end{equation}
for the spherically symmetric metric
\bea
ds^{2} = -N^{2}dt^{2}+\Lambda^{2}(dr+N^{r}dt)^{2}+R^{2}d\Omega^2.
\label{metric}
\eea
The canonical action for this symmetry reduction is given by \cite{kuchavr1994geometrodynamics, husain2005flat}
\begin{equation}
S = \int dtdr\left(P_{\Lambda}\dot{\Lambda}+P_{R}\dot{R}+P_{\phi}\dot{\phi}-NH-N^{r}C_{r}\right),
\label{can_action}
\end{equation}
where $H$ and $C_r$ are the Hamiltonian and radial diffeomorphism constraints. Variation with respect to the phase space variables and the lapse and shift functions $N$ and $N^r$ then leads to the classical equations of motion and the constraints.

As outlined before, for the semiclassical equations, we drop the classical  equations of motion for the scalar field phase space variables $\phi$ and  $P_\phi$ and replace all occurrences of these variables by their expectation values (in an as yet unspecified state) in the remaining equations. This gives the effective constraints
\bea
\langle H\rangle &=& \frac{1}{R^{2}\Lambda}\left(\frac{1}{2}\left(P_{\Lambda}\Lambda\right)^{2}-\left(P_{\Lambda}\Lambda\right)(P_{R}R)\right)\nn\\
&&+\frac{1}{\Lambda^{2}}\left(RR''\Lambda-RR'\Lambda'-\frac{\Lambda^{3}}{2}+\frac{\Lambda R'^{2}}{2}\right)\nn\\
&&+\frac{\langle P_{\phi}^{2}\rangle}{2\Lambda R^{2}}+\frac{R^{2}}{2\Lambda}\langle\phi'^{2}\rangle+\Lambda R^{2}\ \langle V(\phi)\rangle = 0,
\label{Heff}\\\nn\\
\langle C_{r}\rangle &=& P_{R}R'-\Lambda P_{\Lambda}'+\langle P_{\phi}\phi'\rangle = 0
\label{Creff}
\eea
and the effective equations of motion
\bea
\dot{\Lambda} &=& \frac{N}{R^{2}\Lambda}\left(P_{\Lambda}\Lambda^{2}-\Lambda RP_{R}\right)+\left(\Lambda N^{r}\right)',
\label{Lamdot}\\
\dot{P}_{\Lambda} &=& N\left(-\frac{P_{\Lambda}^{2}}{2R^{2}}+\frac{RR''}{\Lambda^{2}}+\frac{1}{2}+\frac{R'^{2}}{2\Lambda^{2}}-\frac{2RR'\Lambda'}{\Lambda^{3}}\right)\nn\\
&&-\left(\frac{RR'N}{\Lambda^{2}}\right)'+\frac{N}{2\Lambda^{2}R^{2}}\left(\langle P_{\phi}^{2}\rangle+R^{4}\langle\phi'^{2}\rangle\right)\nn\\
&&+N^{r}P_{\Lambda}'-NR^{2}\ \langle V(\phi)\rangle,
\label{PLdot}\\
\dot{R} &=& -N\frac{P_{\Lambda}}{R}+N^{r}R',
\label{Rdot}\\
\dot{P}_{R} &=& N\left(\frac{P_{\Lambda}^{2}\Lambda}{R^3}-\frac{P_{R}P_{\Lambda}}{R^{2}}+\frac{\langle P_{\phi}^{2}\rangle}{\Lambda R^{3}}-\frac{R}{\Lambda}\ \langle\phi'^{2}\rangle\right)\nn\\
&&-\left(\frac{R\Lambda'N}{\Lambda^{2}}\right)'+\frac{R'\Lambda'N}{\Lambda^{2}}+\left(\frac{R'N}{\Lambda}\right)'\nn\\
&&-\frac{NR''}{\Lambda}-\left(\frac{NR}{\Lambda}\right)''+\left(N^{r}P_{R}\right)'\nn\\
&&-2N\Lambda R\ \langle V(\phi)\rangle.
\label{PRdot}
\eea
\subsection{Static equations}
The above equations should include the functional Schrodinger equation for the scalar field state, but we do not write it since we consider the six static equations given by
\bea
\dot{\Lambda} &=& \dot{R} = 0,
\label{LRdot}\\
\dot{P}_{\Lambda} &=& \dot{P}_{R} = 0,
\label{Pdots}\\
\langle H\rangle &=& \langle C_{r}\rangle = 0
\label{HCeff}
\eea
with the understanding that the state evolves only by a phase so that the expectation values occurring in (\ref{Pdots})-(\ref{HCeff}) are independent of $t$. This is the static system we study. We work with the gauge choices
\bea
R = r,\quad\Lambda =1
\label{gauge}
\eea
so that the metric (\ref{metric}) reduces to that of the generalized Painleve-Gullstrand (PG) form
\bea
ds^{2} = -N^{2}(r)dt^{2}+ \left(dr+N^{r}(r)\ dt\right)^{2}+r^{2}d\Omega^2.
\label{pgform}
\eea

The six equations (\ref{LRdot})-(\ref{HCeff}) then relate the four gravitational field variables $N, N^{r}, P_{R}$, and $P_{\Lambda}$ with the four scalar field expectation value functions $\langle\phi'^{2}\rangle, \langle P_{\phi}^{2}\rangle, \langle V(\phi)\rangle$, and $\langle P_{\phi}\phi'\rangle$. We ensure that the latter are bounded and satisfy physically relevant fall-off conditions in the coordinate $r$. In particular, since $\langle\phi'^{2}\rangle, \langle P_{\phi}^{2}\rangle$ are expectations of positive definite operators, we require $\langle\phi'^{2}\rangle\geq0$ and $\langle P_{\phi}^{2}\rangle\geq0$ for all $r$. We also impose that $\langle\phi'^{2}\rangle, \langle P_{\phi}^{2}\rangle\,\to\,0$ sufficiently fast at infinity so that matter is confined near $r = 0$. Lastly, the effective diffeomorphism constraint (\ref{Creff}) can be viewed as determining the expectation value $\langle P_{\phi}\phi'\rangle$ after all the other equations are solved for $P_{R}$ and $P_{\Lambda}$ (in the gauges $R = r$ and $\Lambda = 1$); as such, it can be viewed as trivial for the purpose of solving Eqs.~(\ref{LRdot})-(\ref{HCeff}). Thus, there are effectively five equations that relate the seven functions $N, N^{r}, P_{R}, P_{\Lambda}, \langle\phi'^{2}\rangle, \langle P_{\phi}^{2}\rangle$, and $\langle V(\phi)\rangle$.

It remains to simplify the five equations to a form useful for numerical solutions. This may be done in the following steps. In the chosen gauges (\ref{gauge}), Eqs.~(\ref{Lamdot}) and (\ref{Rdot}) can be solved to give
\bea
P_{R} = \frac{\left(rN^{r}\right)'}{N},\quad P_{\Lambda} = \frac{rN^{r}}{N}.
\label{PRPL}
\eea
Substituting these into $\dot{P}_\Lambda = 0$ (\ref{PLdot}) and the effective Hamiltonian constraint (\ref{Heff}), and defining ${\cal N}\equiv N^{2}$ and ${\cal N}^{r}\equiv\left(N^{r}\right)^{2}$, leads to
\bea
{\cal N}' &=& \frac{2\ {\cal N}^{2}\left(r^{4}\langle\phi'^{2}\rangle+\langle P_{\phi}^{2}\rangle\right)}{r^{3}({\cal N}+{\cal N}^{r})},
\label{N2prime}\\\nn\\
{\cal N}^{r\prime} &=& r{\cal N}\left(\langle\phi'^{2}\rangle+2\ \langle V(\phi)\rangle+\frac{\langle P_{\phi}^{2}\rangle}{r^{4}}\right)-\frac{{\cal N}^{r}}{r}.
\label{Nr2prime}
\eea
Lastly, an equation for $\langle V(\phi)\rangle'$ can be derived by substituting $N', N''$, and $\left(N^{r}\right)'$ calculated from (\ref{N2prime}) and (\ref{Nr2prime}) and $P_{R}'$ from (\ref{PRPL}) into Eq.~(\ref{PRdot}), or $\dot{P}_{R} = 0$. This yields an equation of the form
\bea
\langle V(\phi)\rangle' &=& v\left(r, {\cal N}, {\cal N}^{r}, \langle\phi'^{2}\rangle, \langle\phi'^{2}\rangle', \langle P_{\phi}^{2}\rangle, \langle P_{\phi}^{2}\rangle', \langle V(\phi)\rangle\right), \label{Vphi'}\nn\\&&\quad
\eea
where the right-hand side is given in Appendix \ref{app. 1}.

The last three equations are coupled ODEs for ${\cal N}, {\cal N}^{r}$, and $\langle V(\phi) \rangle$ in which $\langle\phi'^{2}\rangle$, $\langle P_{\phi}^{2}\rangle$, and their derivatives may be viewed as sources. That $\langle V(\phi) \rangle$ is also to be determined means that the functional form of the potential is not fixed at the outset but is left arbitrary. In this setting then, a solution of these equations with given sources $\langle\phi'^{2}\rangle$ and $\langle P_{\phi}^{2}\rangle$ determines the metric functions $N$ and $N^{r}$ (and $P_{R}$ and $P_{\Lambda}$ from (\ref{PRPL})). These are the equations we solve numerically for physically motivated sources that correspond to quantum energy density near the origin. This, as we will see, ensures that $\langle V(\phi)\rangle$ tends to a constant at large radius $r$, corresponding to dS, AdS, or asymptotically flat asymptotics.

\section{Static solutions}
\label{sec. 3}
We seek regular static solutions of the equations derived above with boundary data given at $r = 0$ for integration toward large $r$. To find appropriate boundary conditions we assume expansions at $r = 0$ of the form
\bea
N &=& c_{0}+c_{1}r+c_{2}r^{2}+\cdots,\nn\\
N^{r} &=& d_{0}+d_{1}r+d_{2}r^{2}+\cdots,\nn\\
P_{\Lambda} &=& k_{0}+k_{1}r+k_{2}r^{2}+\cdots,\nn\\
P_{R} &=& l_{0}+l_{1}r+l_{2}r^{2}+\cdots,\nn\\
\langle P_{\phi}^{2}\rangle &=& F_{0}+F_{1}r+F_{2}r^{2}+\cdots,\nn\\
\langle\phi'^{2}\rangle &=& G_{0}+G_{1}r+G_{2}r^{2}+\cdots,\nn\\
\langle V(\phi)\rangle &=& V_{0}+V_{1}r+V_{2}r^{2}+\cdots
\eea
and substitute into the expression for the Kretschmann scalar and Eqs.~(\ref{LRdot})-(\ref{HCeff}) to derive conditions on the coefficients. Regularity at $r = 0$ demands finiteness of the Kretschmann scalar there, which implies $k_{0} = k_{1} = 0$. Substituting the lapse and shift expansions into $\dot{\Lambda} = 0$ via (\ref{PRPL}) gives $l_{0} = 0$ and $d_{1}+c_{0}k_{2} = c_{0}l_{1}$, and substituting into $\dot{R} = 0$ via (\ref{PRPL}) gives $d_{0} = 0$ and $d_{1} = c_{0}k_{2}$; hence, $l_{1} = 2k_{2}$ for $c_{0}\ne0$, a condition that is also required for curvature regularity at $r = {0}$. Lastly, substituting the expansions into (\ref{Heff}), (\ref{PLdot}), and (\ref{PRdot}) yields  $F_{0} = F_{1} = F_{2} = F_{3} = 0, c_{1} = 0$, and 
\bea
F_{4}+G_{0}+V_{0} &=& \frac{3}{2}k_{2}^{2},\\
F_{4}+G_{0}-V_{0} &=& -\frac{3}{2}k_{2}^{2}+\frac{2c_{2}}{c_{0}},\\
F_{4}-G_{0}-V_{0} &=& -\frac{3}{2}k_{2}^{2}+\frac{2c_{2}}{c_{0}}.
\eea
These give $F_{4} = c_{2}/c_{0}$, $G_{0} = 0$, and $V_{0} = 3k_{2}^{2}/2-c_{2}/c_{0}$.

In summary, the expansions of the fields at $r=0$, subject to our equations and Kretschmann regularity, are
\bea
N &=& c_{0}+c_{2}r^{2}+\cdots,\nn\\
N^{r} &=& d_{1}r+\cdots,\nn\\
P_{\Lambda} &=& \frac{d_{1}}{c_{0}}r^{2}+\cdots,\nn\\
P_{R} &=& \frac{2d_{1}}{c_{0}}r+\cdots,\nn\\
\langle P_{\phi}^{2}\rangle &=& \frac{c_{2}}{c_{0}}r^{4}+\cdots,\nn\\
\langle\phi'^{2}\rangle &=& G_{1}r+\cdots,\nn\\
\langle V(\phi)\rangle &=& \left(\frac{3}{2}\frac{d_{1}^{2}}{c_{0}^{2}}-\frac{c_{2}}{c_{0}}\right)+V_{1}r+\cdots.
\eea
We note that the leading order term for $\langle V(\phi)\rangle$ is a constant that depends on $c_{0}, c_{2}$, and $d_{1}$. It is  useful to make this constant arbitrary by adding a cosmological constant at the outset. Doing so allows us to write $\langle V(\phi)\rangle = \lambda+V_{1}r+\cdots$, where $\lambda$ is an arbitrary constant. Thus, the boundary conditions for integrating Eqs.~(\ref{N2prime})-(\ref{Vphi'}) are
\bea
{\cal N}(0) = C\ne0,\quad{\cal N}^{r}(0) = 0,\quad\langle V(\phi)\rangle(0) = \lambda
\label{BCs}
\eea
for arbitrary $C$ and $\lambda$. For the source expectation values, we take
\bea
\langle P_{\phi}^{2}\rangle &=& Ar^{4}\exp(-\alpha(r-u_{1})^{2}),
\label{fsource}\\
\langle\phi'^{2}\rangle &=& Br^{2}\exp(-\beta(r-u_{2})^{2}),
\label{gsource}
\eea
where all the parameters are positive real numbers. These are consistent with the boundary conditions on these quantities at $r = 0$, and they represent ``lumps" of scalar field peaked close to the origin. The Gaussian fall-off here implies that Eqs.~(\ref{N2prime})-(\ref{Vphi'}) for large radial values simplify to
\bea
{\cal N}' &=& 0,\\
{\cal N}^{r\prime} &=& 2r{\cal N}\langle V(\phi)\rangle-\frac{{\cal N}^{r}}{r},\\
\langle V(\phi)\rangle' &=& 0.
\label{largeR}
\eea
These have solutions $N = \text{constant}$ and $\langle V(\phi)\rangle = \text{constant}$ as $r\rightarrow\infty$; the corresponding solution for the asymptotic shift is $N^{r} = \sqrt{a r^{2}+b/r}$ for constants $a, b$; as we will see, $a$ may be positive, negative, or zero as it denotes the cosmological constant, and $b\ge0$ since it represents the mass parameter. It is evident that for the AdS case, where $a<0$, this coordinate system breaks down for sufficiently large $r$ \cite{faraoni2020painleve}, but it is nevertheless sufficient to establish the type of solutions that can follow from Eqs.~(\ref{N2prime})-(\ref{Vphi'}).

In the following, we exhibit boson star and black hole solutions of these equations with the source terms (\ref{fsource})-(\ref{gsource}); these are characterized by the parameters $A, B, \alpha, \beta, u_{1}, u_{2}$ and the boundary conditions (\ref{BCs}). To distinguish boson star and black hole solutions, we use the condition for the apparent horizon, which for the metric (\ref{pgform}) is
\bea
|\nabla_{a}r|^{2} = 1-\frac{{\cal N}^{r}}{{\cal N}} = 0.
\label{ah}
\eea
We plot this function together with the scalar field energy density
\bea
\rho = \frac{\langle P_{\phi}^{2}\rangle}{2r^{4}}+\frac{1}{2}\langle\phi'^{2}\rangle,
\label{rhophi}
\eea
$\langle V(\phi)\rangle$, and the Kretschmann scalar.

\subsection{Boson stars}
\label{bosonstars}
Boson stars are interesting equilibrium solutions to Einstein-Klein-Gordon systems, where equilibrium is attained through a counterbalance between gravitational attraction and the dispersive nature of scalar fields \cite{jetzer1992boson}. Although initially found as classical solutions, there have also been attempts to study them in semiclassical gravity \cite{ho2002boson, bernal2010multistate, alcubierre2023boson}. All such attempts involve the calculation of $\langle\hat{T}_{ab}\rangle$ and the assumptions therein. Our approach avoids this calculation and takes into account the full backreaction of quantum fields on classical geometry.

\begin{figure}
\centering
\includegraphics[width = \linewidth]{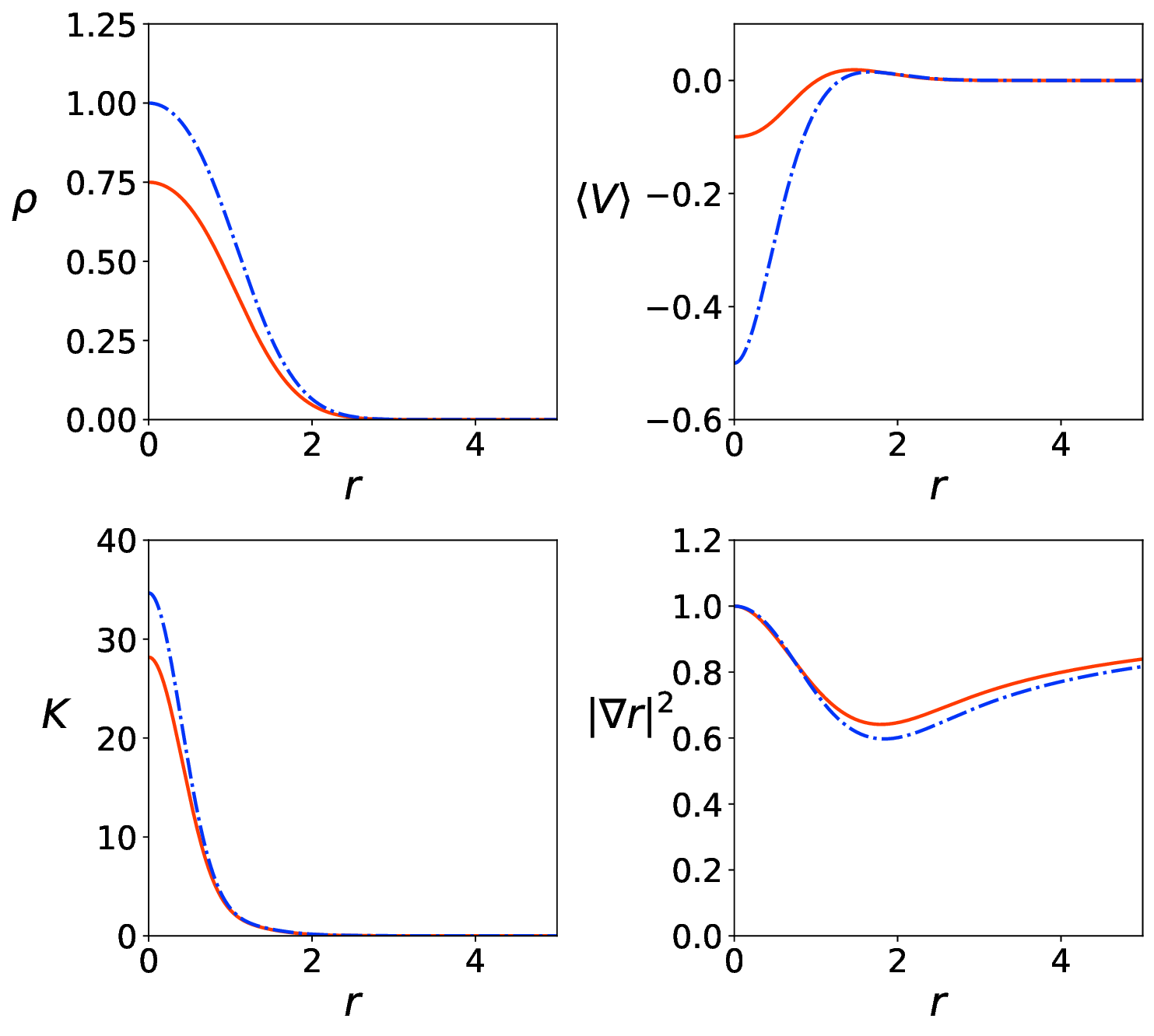}
\caption{\textbf{Asymptotically flat boson stars:} $\rho$ and $\langle V\rangle$ are the expectation values of the scalar field density and potential; $K$ is the Kretschmann scalar; and the apparent horizon function $|\nabla r|^2$ has no roots. (The red curves are for parameter values $A = 1.5, \alpha = 1, u_{1} = 0, B\approx0.898, \beta = 1, u_{2} = 0$ and boundary data ${\cal N}(0) = 1, {\cal N}^{r}(0) = 0, \langle V(\phi)\rangle(0) = -0.1$; the blue ones are for $A = 2, \alpha = 1, u_{1} = 0, B\approx1.28, \beta = 1, u_{2} = 0$ and ${\cal N}(0) = 1, {\cal N}^{r}(0) = 0, \langle V(\phi)\rangle(0) = -0.5$.)}
\label{Fig:Bstar}
\end{figure}

Fig. \ref{Fig:Bstar} exhibits sample asymptotically flat boson star solutions obtained by integrating Eqs.~(\ref{N2prime})-(\ref{Vphi'}) from $r=10^{-7}$ outward. As mentioned above, for nearly all values of the parameters in (\ref{fsource})-(\ref{gsource}) and boundary data, $\langle V(\phi)\rangle$ tends to a nonzero constant for large $r$, indicating that such solutions are dS or AdS boson stars. One way of obtaining $\langle V(\phi)\rangle = 0$ as $r$ increases requires fine-tuning of one of the parameters. Fig. \ref{Fig:Bstar} exhibits one such solution obtained by tuning the amplitude $B$ in (\ref{gsource}) to $B\approx0.898$, obtained using a numerical binary search. The first frame in Fig. \ref{Fig:Bstar} is the source energy density (\ref{rhophi}); the second frame shows $\langle V(\phi)\rangle$, which tends to zero as $r$ increases, indicating asymptotic flatness---the minimum in this function near $r = 0$ may be an indication of stability; the third frame is the Kretschmann scalar; and the last is the apparent horizon function (\ref{ah}), which has no roots, hence no horizons. Our solutions are qualitatively similar to those in Ref.~\cite{alcubierre2023boson}, which obtains boson star solutions by calculating $\langle\hat{T}_{ab}\rangle$ in the SSCs, or semiclassical self-consistent configurations, approach. Values of $B$ other than the tuned value indicated give dS and AdS boson star solutions.

\subsection{Regular asymptotically flat black holes}
\label{Sec:bh1}
\begin{figure}
\centering
\includegraphics[width = \linewidth]{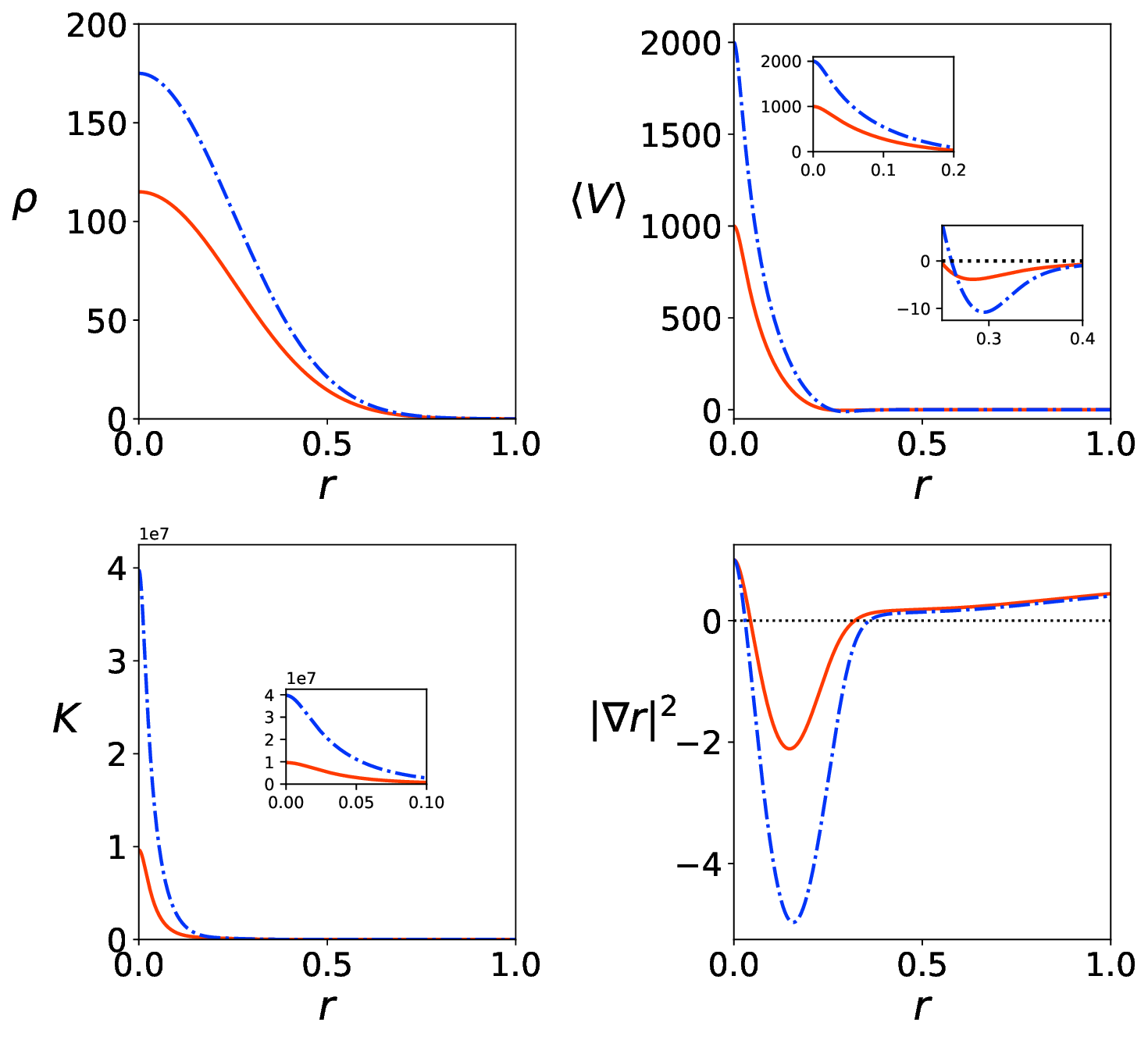}
\caption{\textbf{Asymptotically flat black holes:} $\rho$ and $\langle V\rangle$ are the expectation values of the scalar field density and potential; $K$ is the Kretschmann scalar; $|\nabla r|^2$ shows an inner horizon at $r\approx0.1$ and an outer one at $r\approx0.4$. (The red curves are for the parameters $A = 230, \alpha = 10, u_{1} = 0, B\approx498.663, \beta = 10, u_{2} = 0$ and boundary data ${\cal N}(0) = (0.02)^{2}, {\cal N}^{r}(0) = 0, \langle V(\phi)\rangle(0) = 1000$; the blue ones are for $A = 350, \alpha = 10, u_{1} = 0, B\approx659.851, \beta = 10, u_{2} = 0$ and ${\cal N}(0) = (0.01)^{2}, {\cal N}^{r}(0) = 0, \langle V(\phi)\rangle(0) = 2000$.}
\label{Fig:bh1}
\end{figure}

The signature of a black hole solution is the presence of a trapped region as indicated by roots of the function $|\nabla r|^{2}$ (\ref{ah}). That there are solutions of Eqs.~(\ref{N2prime})-(\ref{Vphi'}) with this property is shown in Fig. \ref{Fig:bh1} for the two sets of parameter values indicated in the figure caption. These parameter values are obtained by starting from boson star solutions, increasing the amplitudes $A$ and $B$ (see (\ref{fsource})-(\ref{gsource})), and changing the boundary conditions until the appearance of the roots of $|\nabla r|^{2}$. With all other solution parameters prescribed, the parameter $B$ may again be tuned via binary search to obtain asymptotically flat solutions, this method being one of numerous possibilities for parameter searches. Several features of the asymptotically flat black hole solutions in Fig. \ref{Fig:bh1} are noteworthy: with the tuning of $B$, $\langle V(\phi) \rangle$ rapidly falls to zero as shown in the top right frame---the inset figures show a close-up of this function indicating minima; the boundary value of the lapse function at $r = 0$ is $N\approx 0.02$, which results in a large ($\sim10^{7}$) but finite value of the Kretschmann scalar at $r = 0$ (as detailed in the close-up inset frame); for the two sets of parameters, there is an outer horizon near $r \approx 0.4$ and an inner horizon closer to the origin; the matter density and potential (upper row) are both bounded; just as for boson stars, the dip in the potential $\langle V(\phi)\rangle$ may again hint at stability although establishing it would require a separate analysis; and Fig. \ref{Fig:bh1} indicates that the expectation value of the density shows ``leakage" outside the outer horizon (located near $r\approx 0.4$), which may be taken as an indication of ``quantum hair."

\subsection{dS and AdS black holes}
\label{Sec:bh2}
\begin{figure}
\centering
\includegraphics[width = \linewidth]{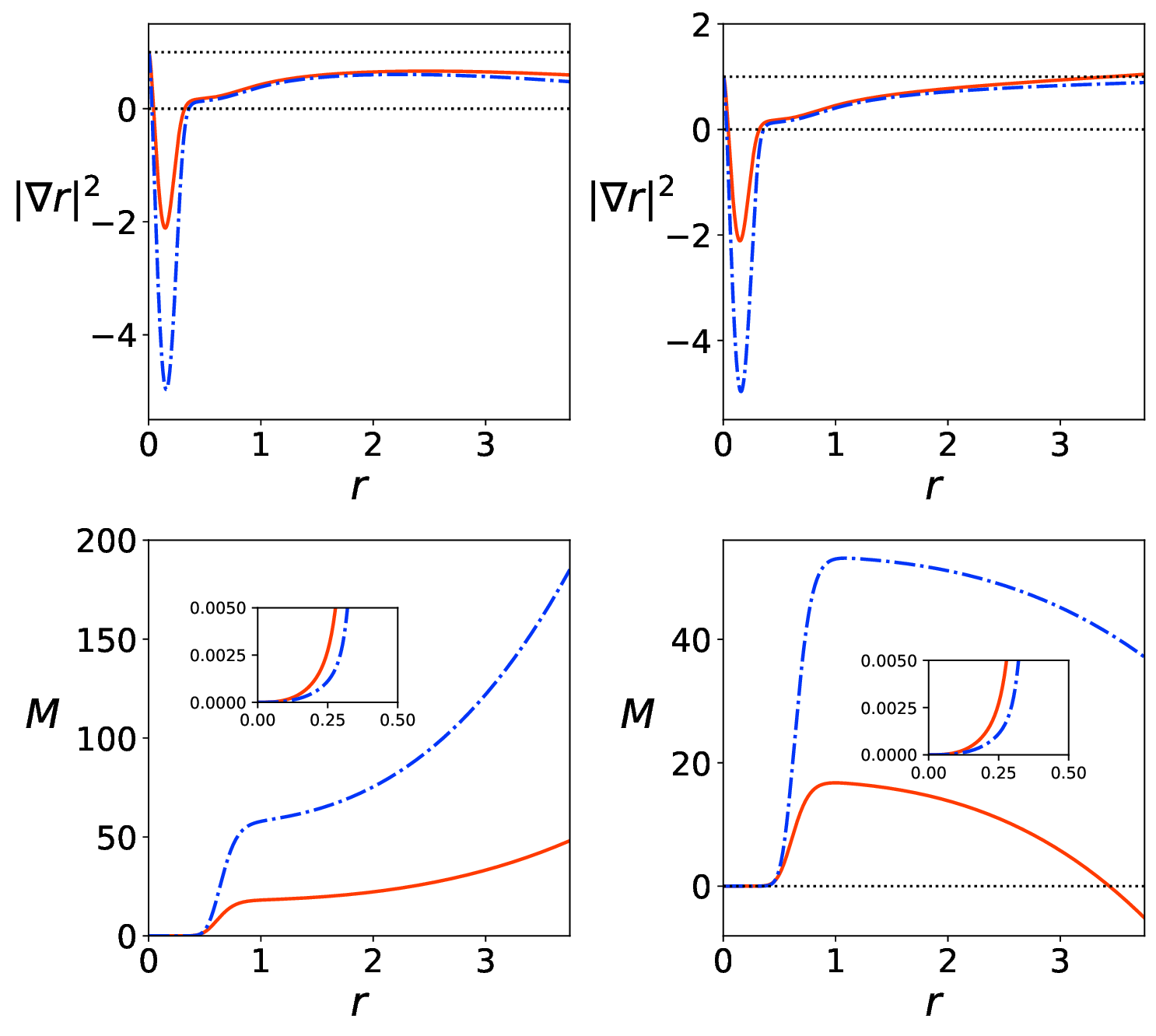}
\caption{\textbf{Regular dS and AdS black holes:} the upper row shows the apparent horizon function with roots at the outer and inner horizons; the lower row shows the mass function with cosmological constant (\ref{Mfn}). (All data are the same as those used for Fig. \ref{Fig:bh1} except that $B = 503.5$ (red) and $B = 670$ (blue) for the dS case and that $B = 495$ (red) and $B = 658.5$ (blue) for the AdS case.)}
\label{Fig:bh2}
\end{figure}
We noted above that obtaining asymptotically flat black hole solutions required fixing boundary conditions and other parameters such that $\langle V(\phi)\rangle\rightarrow0$ for large $r$ and that one way to do so was by fine-tuning $B$ in (\ref{gsource}). Obtaining regular dS and AdS black holes and boson stars is much easier computationally as this requires, for instance, changing the value of $B$ from that used for the asymptotically flat example above. Fig. \ref{Fig:bh2} shows sample solutions obtained in this way that are dS (left column) and AdS (right column); the top row is the apparent horizon function $|\nabla r|^2$ that has two roots for the inner and outer horizons; and the second row shows the mass function $M(r)$ defined by
\bea
{\cal N}^{r}\equiv 2M(r)/r = 2m(r)/r+\Lambda r^{2}/3,
\label{Mfn}
\eea
where $m(r)$ is the mass function associated to the scalar field. As $r$ increases, $M(r)\sim\pm r^3$ as expected for the dS and AdS cases.

In summary, in this section, we have shown numerically that our proposed canonical semiclassical equations (\ref{N2prime})-(\ref{Vphi'}) have regular boson star and black hole solutions that are asymptotically flat, dS, or AdS.

\section{Comparison with other regular black hole proposals}
\label{sec. 4}
Among the proposals for nonsingular black holes in the literature, the most prominent are the Hayward and Bardeen metrics and their generalization \cite{frolov2016notes}; the metric is of the form  
\bea
ds^{2} = -fA^{2}dv^{2}+2Advdr+r^{2}d\Omega^{2},
\label{hay}
\eea
where $f(r)$ and $A(r)$ are chosen for regularity of the curvature at $r = 0$ and asymptotic flatness. A comparison of this proposal with the semiclassical solutions we find proceeds by considering Eqs.~(\ref{LRdot})-(\ref{HCeff}) with the shift function $N^{r} = 0$ and transformation to the Eddington-Finkelstein (EF) coordinates $dt = dv-\left(\Lambda/N\right)dr$ followed by a comparison to the solutions with the functions $f(r), A(r)$ in \cite{frolov2016notes}.

With $N^{r} = 0$ and $R = r$, the metric (\ref{metric}) transformed to EF coordinates and compared with (\ref{hay}) leads to
\bea
f = \Lambda^{-2},\quad A = N\Lambda.
\label{fA}
\eea
Our equations for $\Lambda$ and $N$ derived from (\ref{Heff})-(\ref{PRdot}) are
\bea
(N^{2})' &=& rN^{2}\left(\langle\phi'^{2}\rangle+\frac{\langle P_{\phi}^{2}\rangle}{r^{4}}-2\Lambda^{2}\langle V(\phi)\rangle+\frac{\Lambda^{2}-1}{r^{2}}\right),\nn\\
\label{n2haybar}
\\
(\Lambda^{2})' &=& rN^{2}\left(\langle\phi'^{2}\rangle+\frac{\langle P_{\phi}^{2}\rangle}{r^{4}}+2\Lambda^{2}\langle V(\phi)\rangle-\frac{\Lambda^{2}-1}{r^{2}}\right).\nn\\
\label{l2haybar}
\eea
Substituting (\ref{fA}) into these equations with $A = 1$ as in the Hayward and Bardeen metrics gives, respectively,
\bea
f' &=& r\left[f\left(\frac{\langle P_{\phi}^{2}\rangle}{r^{4}}+\langle\phi'^{2}\rangle\right)-2\langle V(\phi)\rangle+\frac{(1-f)}{r^{2}}\right],
\label{subn2haybar}
\nn\\
-f' &=& r\left[f\left(\frac{\langle P_{\phi}^{2}\rangle}{r^{4}}+\langle\phi'^{2}\rangle\right)+2\langle V(\phi)\rangle+\frac{(f-1)}{r^{2}}\right].\nn\\
\label{subl2haybar}
\eea
Adding these gives 
\bea
\langle P_{\phi}^{2}\rangle = -r^{4}\langle\phi'^{2}\rangle,
\eea
which shows that the condition required for positivity of both the expectation values $\langle\phi'^{2}\rangle, \langle P_{\phi}^{2}\rangle\geq0$ is not satisfied. Hence, the Bardeen and Hayward metrics do not arise from our semiclassical scalar field equations. This is the case for any metric of the form (\ref{hay}) with $A = 1$.

There are nonsingular metric proposals for $A\ne1$ reported in \cite{frolov2016notes}, where the mass $M$ of a black hole is a function of the radii of its inner and outer horizons. It is readily verified that this relation is not satisfied for the sample asymptotically flat solution in Fig. \ref{Fig:bh1} where the inner horizon is at $\sim0.1$, the outer horizon is at $\sim0.4$, and the black hole mass is $\sim15$ as determined by (\ref{Mfn}) with $\Lambda = 0$; the same holds for the second case in Fig. \ref{Fig:bh1}. These considerations show that the semiclassical solutions we present are not the same as the regular black hole metric proposals in the literature.

\section{Summary and discussion}
\label{sec. 5}

We have shown that the Hamiltonian semiclassical proposal defined by (\ref{static1}) for the case of the scalar field in spherical symmetry admits asymtotically flat, dS, and AdS black holes and boson stars. We provide sample numerical solutions for each of these classes.

The method we used is significantly different from all previous approaches to semiclassical gravity in that we do not begin with a particular quantum state defined on a background and compute the expectation value of the stress energy tensor. We leave the state unspecified but fixed and focus instead on the expectation values of the matter operators that appear in the equations---with the restriction that these satisfy positivity properties. As we argue, the expectation values $\langle\phi'^{2}\rangle, \langle P_{\phi}^2\rangle$ could be viewed as independent sources in the semiclassical equations whereas $\langle V(\phi)\rangle$ and the lapse and shift functions are determined by Eqs.~(\ref{N2prime})-(\ref{Vphi'}). The latter means that the scalar theory is specified only up to fixing the potential and the expectation value of the potential in the fixed state is determined by the equations. The expectation values of the sources are chosen to correspond to the physical intuition of matter confined near the origin, and the solutions are specified by the parameters therein.

Of particular note in the solutions is that the ``tails" of the expectation values of the matter density (\ref{rhophi}) in Figs. \ref{Fig:bh1} and \ref{Fig:bh2} extend beyond the horizon. This may be viewed as ``quantum hair." It is similar to classical hair, which may be defined as one or more parameters in a black hole solution that are not captured by conserved integrals at infinity \cite{david1997black}. It is possible to take sources of compact support rather than the choices (\ref{fsource}) and (\ref{gsource}), but the point here is to establish that solutions with tails are also valid and allowed.

For comparison with other works, we note that the regular black hole solutions presented here have not been obtained in the context of the covariant semiclassical Einstein equation (\ref{scee}). Instead, there are results showing that a wormhole-like geometry replaces the classical black hole \cite{berthiere2018fate, kain2025quantum} and that a rotating black hole arises from using the Weyl anomaly as the source \cite{fernandes2023rotating}.

Wormholes \cite{morris1988wormholes,visser1995lorentzian} are another interesting class of metrics that we did not report for the purposes of this paper. These have been studied using the covariant approach to the semiclassical Einstein equation \cite{hochberg1997self}. Following the ansatz in \cite{morris1988wormholes}, it is possible to consider our equations in the gauge $R = \sqrt{r^{2}+k^{2}}$ (instead of $R=r$ as above); the result is a set of three coupled equations similar to (\ref{N2prime})-(\ref{Vphi'}) with the additional adjustable parameter $k$. However, the solutions that result have the lapse function $N$ going to zero at $r = 0$, giving a degenerate but nonsingular metric, a feature that may be due to our choice of the generalized PG coordinates. Work on these solutions continues and would be reported elsewhere. Lastly, our equations also give asymptotically flat solutions that are naked singularities, similar to but different from the classical Wyman solution \cite{wyman1981static} for the massless, minimally coupled scalar field. This shows that semiclassical gravity, at least in the form we have studied, does not in general resolve curvature singularities.

There are at least two possible directions for future work: the stationary axisymmetric case and the dynamical solutions in spherical symmetry. The latter in particular would be of interest in the study of semiclassical gravitational collapse.
\medskip

\noindent{\bf Acknowledgements:} This work was supported by the Natural Science and Engineering Research Council (NSERC) of Canada.

\appendix

\section{Differential equation for the expectation of potential energy density}
\label{app. 1}
The differential equation for $\langle V(\phi)\rangle$ is given by
\begin{equation}
\langle V(\phi)\rangle' = \frac{E_{1}}{E_{2}},
\label{eq. a1}
\end{equation}
\newpage
where
\begin{equation*}
\begin{aligned}[b]
E_{1} =&\:2\langle P_{\phi}^{2}\rangle^{2}{\cal N}^{2}{\cal N}^{r}+4{\cal N}^{3}\langle\phi'^{2}\rangle r^{6}\\
&\:-4{\cal N}^{3}\langle P_{\phi}^{2}\rangle r^2-\langle P_{\phi}^{2}\rangle^{2}{\cal N}^{3}\\
&\:-8{\cal N}^{2}{\cal N}^{r}\langle\phi'^{2}\rangle\langle V(\phi)\rangle r^{8}-2{\cal N}({\cal N}^{r})^{2}\langle\phi'^{2}\rangle\langle V(\phi)\rangle r^{8}\\
&\:-8{\cal N}^{2}{\cal N}^{r}\langle P_{\phi}^{2}\rangle\langle V(\phi)\rangle r^{4}-2{\cal N}({\cal N}^{r})^{2}\langle P_{\phi}^{2}\rangle\langle\phi'^{2}\rangle r^{4}\\
&\:-2{\cal N}({\cal N}^{r})^{2}\langle P_{\phi}^{2}\rangle\langle V(\phi)\rangle r^{4}+4\langle\phi'^{2}\rangle\langle P_{\phi}^{2}\rangle{\cal N}^{2}{\cal N}^{r}r^{4}\\
&\:-({\cal N}^{r})^{3}\langle\phi'^{2}\rangle'r^{7}+{\cal N}^{3}\langle\phi'^{2}\rangle'r^{7}\\
&\:+({\cal N}^{r})^{3}\langle\phi'^{2}\rangle r^{6}+{\cal N}^{3}\langle P_{\phi}^{2}\rangle'r^{3}\\
&\:-({\cal N}^{r})^{3}\langle P_{\phi}^{2}\rangle'r^{3}+({\cal N}^{r})^{3}\langle P_{\phi}^{2}\rangle r^{2}\\
&\:-{\cal N}({\cal N}^{r})^{2}\langle P_{\phi}^{2}\rangle^{2}-{\cal N}({\cal N}^{r})^{2}\langle\phi'^{2}\rangle^{2}r^{8}\\
&\:-\langle\phi'^{2}\rangle^{2}{\cal N}^{3}r^{8}+\langle\phi'^{2}\rangle'{\cal N}^{2}{\cal N}^{r}r^{7}\\
&\:-\langle\phi'^{2}\rangle'{\cal N}({\cal N}^{r})^{2}r^{7}+\langle P_{\phi}^{2}\rangle'{\cal N}^{2}{\cal N}^{r}r^{3}\\
&\:-\langle P_{\phi}^{2}\rangle'{\cal N}({\cal N}^{r})^{2}r^{3}+11\langle\phi'^{2}\rangle{\cal N}^{2}{\cal N}^{r}r^{6}\\
&\:+8\langle\phi'^{2}\rangle {\cal N}({\cal N}^{r})^{2}r^{6}-2\langle\phi'^{2}\rangle\langle P_{\phi}^{2}\rangle {\cal N}^{3}r^{4}\\
&\:-6\langle V(\phi)\rangle\langle P_{\phi}^{2}\rangle{\cal N}^{3}r^{4}-5\langle P_{\phi}^{2}\rangle{\cal N}^{2}{\cal N}^{r}r^{2}\\
&\:+2\langle\phi'^{2}\rangle^{2}{\cal N}^{2}{\cal N}^{r}r^{8}-6\langle\phi'^{2}\rangle\langle V(\phi)\rangle{\cal N}^{3}r^{8}
\end{aligned}
\end{equation*}
and
\begin{equation*}
E_{2} = 2r^7({\cal N}^{3}+3{\cal N}^{r}{\cal N}^{2}+3({\cal N}^{r})^{2}{\cal N}+({\cal N}^{r})^{3}).
\end{equation*}

\bibliography{ref}
\end{document}